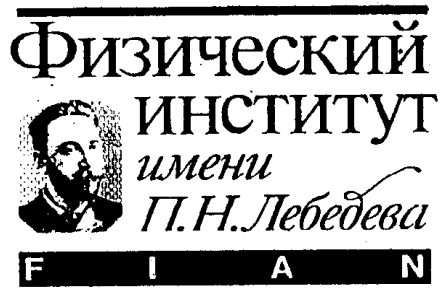

M.A. NEGODAEV, B.N. LOMONOSOV, S.K. KOTELNIKOV,
A.V. BAGULYA, E.M. NEGODAEVA, A.N. LARITCHEV,
V.YU. ZHUKOV, L.N. SMIRNOVA

# GAS ELECTRON MULTIPLIER: PERFORMANCE AND POSSIBILITIES

# GAS ELECTRON MULTIPLIER: PERFORMANCE AND POSSIBILITIES

Negodaev M.A.[*], Lomonosov B.N., Kotelnikov S.K.,

Bagulya A.V., Negodaeva E.M.

*Lebedev Physical Institute, Moscow, Russia*

Laritchev A.N., Zhukov V.Yu., Smirnova L.N.

*Institute of Nuclear Physics, Moscow State University, Russia*

## ABSTRACT

The gas electron multiplier produced with the polyimide film 100 µm thick. The maximum gas gain obtained with gas mixture Ar/CH4(91/9) was 5000. No decrease of gas gain with rates up to $2 \cdot 10^4$ $1/(mm^2 \cdot s)$ was observed. The possibility of operation of GEM with and without MSGC is discussed.

[*] Corresponding author. E-mail: negodaev@sgi.lebedev.ru, negod@sci.lebedev.ru

From the end of 80-s gas position sensitive detectors has been developed with use of microelectronics technologies involved in production of gas amplification structures. As a result of this development the Micro Strip Gas Chambers (MSGC) [1], Micro Gap detector [2], and Micro Mesh Gas detector (MICROMEGAS) [3] were created. The specific feature of all of these detectors is the separation if the drift area of primary ionization electrons and that of gas amplification, with the electrode spacing been 10–100 μm. This enables the detector to operate under high rate conditions up to $10^6$ p/(mm$^2$·s). The position accuracy is determined than by electron diffusion in the gas and usually is about 40 μm [4]. The electrodes of the amplification cell are produced by means of litography on the dielectric support. Presence of the dielectric in the electrode area changes the electric field and as a rule causes the decrease of gas gain in comparison with wire chambers. So the gain in the MSGC with strip pitch of 200 μm does not exceed $10^4$ with the energy resolution been about 17 % [5]. Besides the dielectric presence leads to the amplification dependence on the rate due to charging up effect. The maximum gain value increases following the improvement of accuracy of the amplification structures production with the cost inevitably increases.

To increase gain in the MSGC [6] the Gas Electron Multiplier as an additional amplification element of MSGC layout was offered. First test results obtained with GEM, produced in Moscow are presented here.

GEM consists of polyimide film 100 μm in thickness with copper layers 18 μm thick covering it over both sides. Holes 60 μm in diameter are made by means of litography in the film (fig. 1). The operation area size of the detector tested was 3x3 cm$^2$. GEMs with sensitive area 10x10 cm$^2$ are also produced. The detailed description of manufacturing technology one can find in [7].

GEM was placed 3 mm over the MSGC microstrip pattern. Drift electrode of graphite covered 100 μm polyimide film was placed 3 mm over GEM. The microstrip electrodes pattern of MSGC was produced of aluminium over the borosilicate glass D-263(DESAG, Germany) support 0.3 mm thick. The pattern pitch was 200 μm with anode and cathode width being correspondingly 9 and 90 μm. Detector was installed in hermetic box and gas mixture $Ar/CH_4$ (91/9) was flown through.

The electronic scheme of the set up is shown on fig. 2. The voltage $V_{g1}$ and $V_{g2}$ difference defined the GEM voltage $V_g$. The signal was read out from 16 anode strip joined together. To measure gas gain the integrating amplifier was used with $\tau_u$ = 500 μs, transmission coefficient $2mV/10^4$ $e^-$ and noise with the detector inserted 2000 $e^-$. The detector was irradiated with X-rays with the γ energy E = 8 keV. Fig. 3 shows the amplitude distribution, obtained with single operated MSGC (1) as well as with concurrent operating MSGC + GEM (2).

In measuring GEM gain voltage on MSGC varied to make signal fit in the dynamic range of the amplifier. The MSGC gain versus cathode strips voltage $V_c$ is shown on fig. 4.

Fig. 5 shows the plotted GEM gain versus $V_g$ voltage. During measurements the drift electrode voltage $V_d$ was 100 V over the $V_{g2}$, with $V_{g1}$ being 1400 V. With GEM voltage $V_g$ = 540 V gain obtained was $M_g$ = 1020.

It is necessary to notice that we measured not pure gas gain, but the effective one, defined by the amount of secondary electrons from GEM, reached the MSGC support, that is product of gain and GEM «transparency». The transparency depends on GEM voltage $V_g$ and on ratio between electric fields over GEM film $E_u$ and under it $E_d$.

Fig. 6 shows signal amplitude versus drift field $E_{d2}$ with fixed field $E_{d1}$ = 3.35 kV/cm. Even when $E_{d1}$ is very high and all primary ionization electrons are collected some fractions of electrons produced in process of gas amplification in GEM can reach GEM down electrode. Thus the gas gain of GEM can be underestimated. We expect to study this effect in further investigations.

To study GEM rate capability the γ-source was collimated to form a spot with 1.4 mm$^2$ area in GEM plane. To register the MSGC signal we used fast amplifier with charge and discharge time correspondingly 10 and 100 ns. Fig. 7 shows the signal amplitude versus primary ionization intensity. Specific increase of gain of MSGC on boro-silicate D-263 glass with increase of rate up to $10^3$ γ/(mm$^2$·s) observed earlier [6] and is connected with charging up of the support surface. The rate further increases, the decrease to 60% replaces this gain increase, when the rate reaches value of $2·10^4$ γ/(mm$^2$·s). Analogues behavior of gain takes place in case of GEM.

The gain growth with the rate increase in GEM can be connected with polyimide charging up in holes. This leads to increase of field gradient inside of holes and in turn to that of gain. When the rate is as high as $2·10^4$ γ/(mm$^2$·s) the GEM gain increases on 70 % more with the stable detector operation. Further increase of rate was limited by gamma source possibilities.

Rather high GEM gain allows to use this device without MSGC. An ordinary printed plate with printed strip pattern was used as a readout system instead of MSGC.

The dependence the signal amplitude on the field tense $E_{d1}$ is shown on the fig. 8. The amplitude distribution obtained with the voltage $V_g$ = 568 V is shown on fig. 9. The effective gas gain with this voltage applied was about 5000.

The first results obtained enable to consider the production of a cheap position sensitive detector (the cost of MSGC is 5–10 times of that of GEM). The operation of GEM with the read out strips placed close to GEM or etched on the GEM directly (on one or on both sides) seems very promising. This will allow to increase the life time of the detector.

If the short circuit appears in an ordinary GEM structure this leads to detector failure, but in case of the GEM film with strip pattern a failure of one channel will not affect the operation of a detector as a whole.

The detector can be applied as a track detector in the accelerator set ups as well as in tomography and structure analysis.

# REFERENCES

[1] Oed A. Nucl. Instr. Meth., **A263**, 351 (1988).

[2] Angelini F., Bellazini R., Brez A. et al. Nucl. Instr. Meth., **A335**, 69 (1993).

[3] Giomataris Y., Rebourgeard P., Charpak P. et al. Nucl. Instr. Meth., **A376(1)**, 29 (1996).

[4] Vandervelde C., Bouhali O., Vandoninck W. Nucl. Instr. and Meth., **A378**, 432 (1996).

[5] Berg F.D., Udo F., Zhukov V. et al. Nucl. Instr. Meth., **A401**, 156 (1997).

[6] Sauli F. Nucl. Instr. Meth., **A386**, 531 (1997).

[7] M.A. Negodaev et al. Preprint FIAN №17, 1998.

FIGURE CAPTIONS

Fig. 1. GEM layout. The metallization area is filled with dashed line.

Fig. 2. The detector electronic scheme: 1 — γ-source; 2 — collimator; 3 — mylar window; 4 — drift electrode; 5 — GEM; 6 — MSGC; 7 — 16 joined anode strips.

Fig. 3. The amplitude distribution, obtained in operation of single MSGC (1) and MSGC + GEM (2): $V_c$ = 500 V, $V_{g1}$ = 1400 V, $V_{g2}$ = 1700 V, $V_d$ = 1800 V.

Fig. 4. MSGC gain versus cathode strip voltage $V_c$.

Fig. 5. GEM gain versus $V_g$ voltage: $E_u / E_d$ = 0.1.

Fig. 6. Signal amplitude versus drift field $E_{d2}$ with fixed $E_{d1}$ = 3.35 kV/cm: $V_{g1}$ = 1.4 kV, $V_c$ = 400 V.

Fig. 7. Signal amplitude versus primary ionization intensity: ▲ — MSGC ($M$ = 800), $V_d$ = 1400 V; ● — GEM ( $M$ = 80 ) + MSGC ( $M$ = 9), $E_u / E_d$ = 0.1.

Fig. 8. Signal amplitude versus field tense $E_{d1}$: $V_g$ = 570 V.

Fig. 9. Amplitude distribution obtained with $V_g$ = 568 V: $V_{g1}$ = 506 V , $V_{g2}$ = 1074 V, $V_d$ = 1200 V.

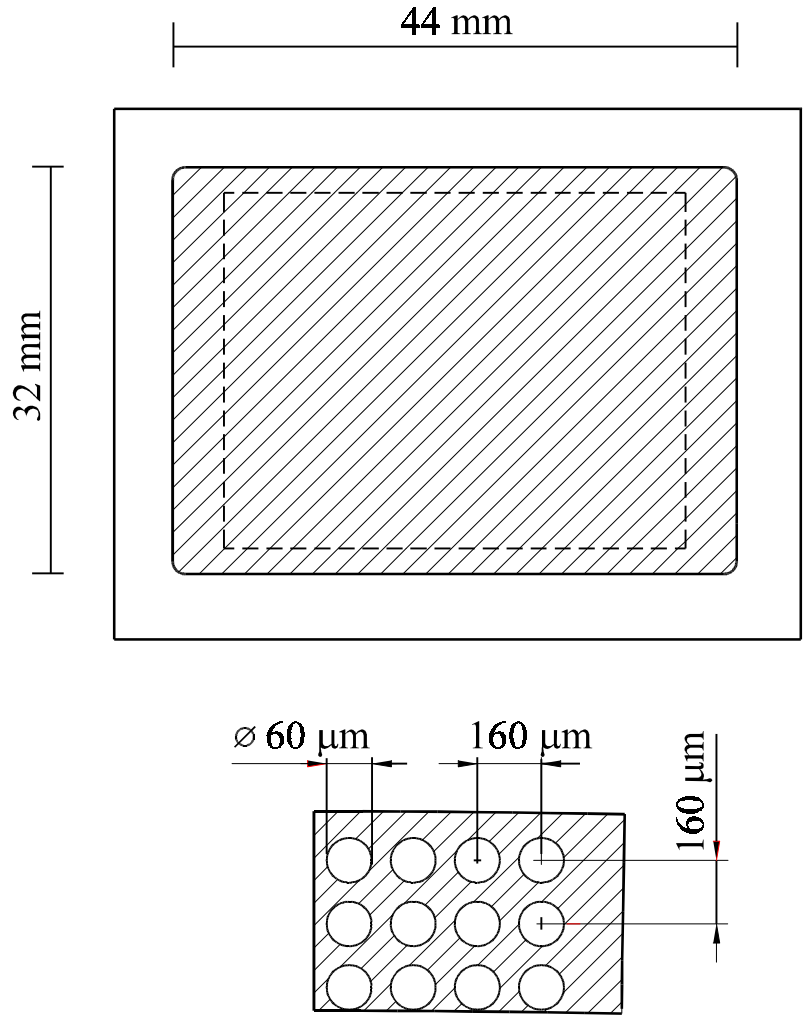

Fig. 1

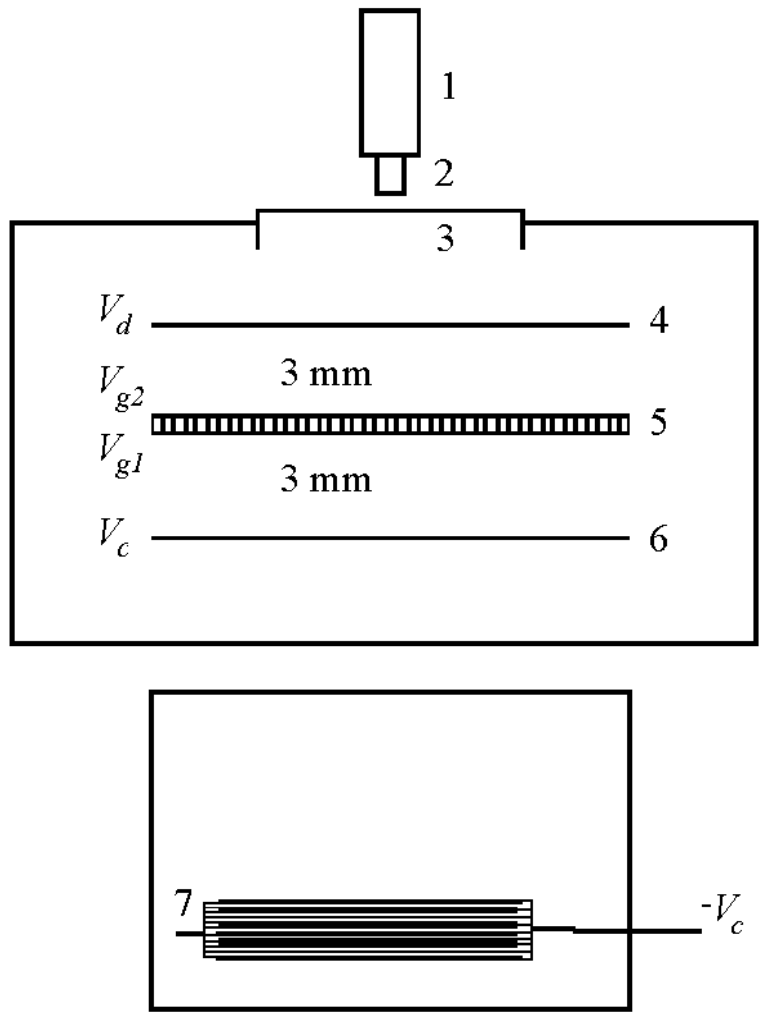

Fig. 2

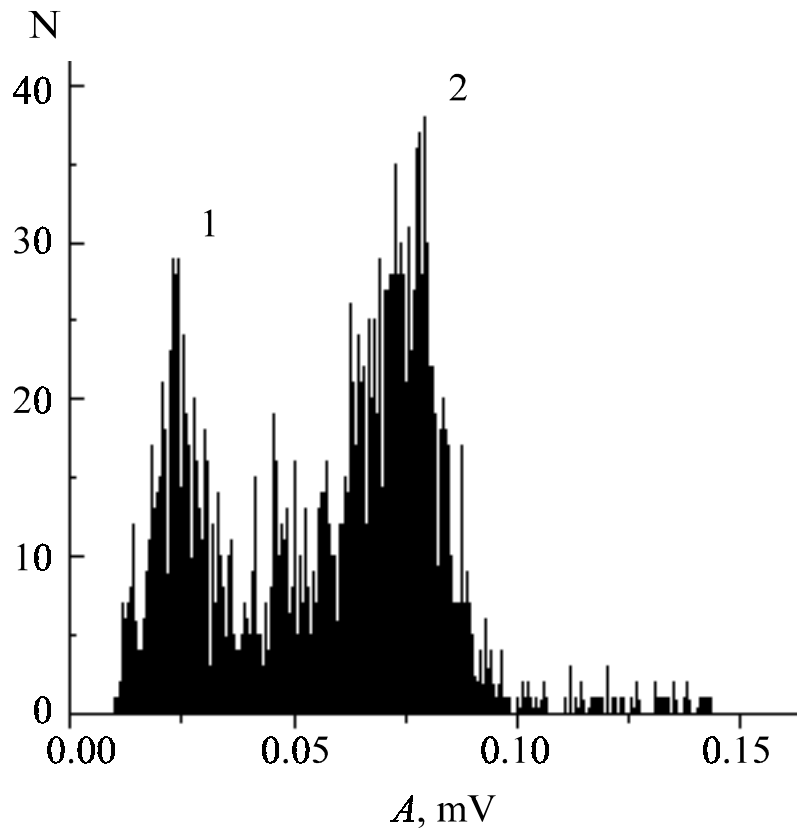

Fig. 3

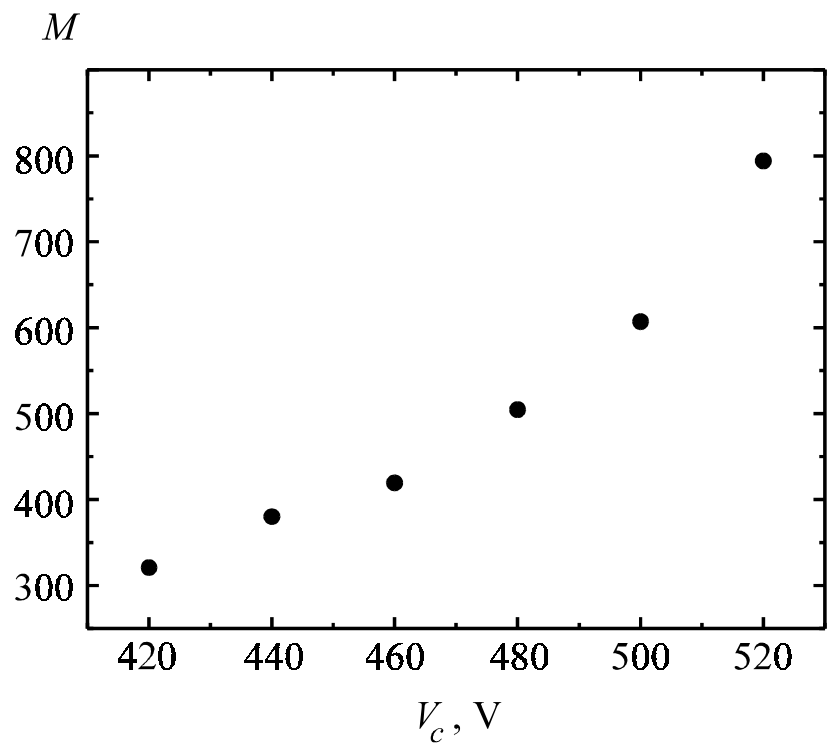

Fig. 4

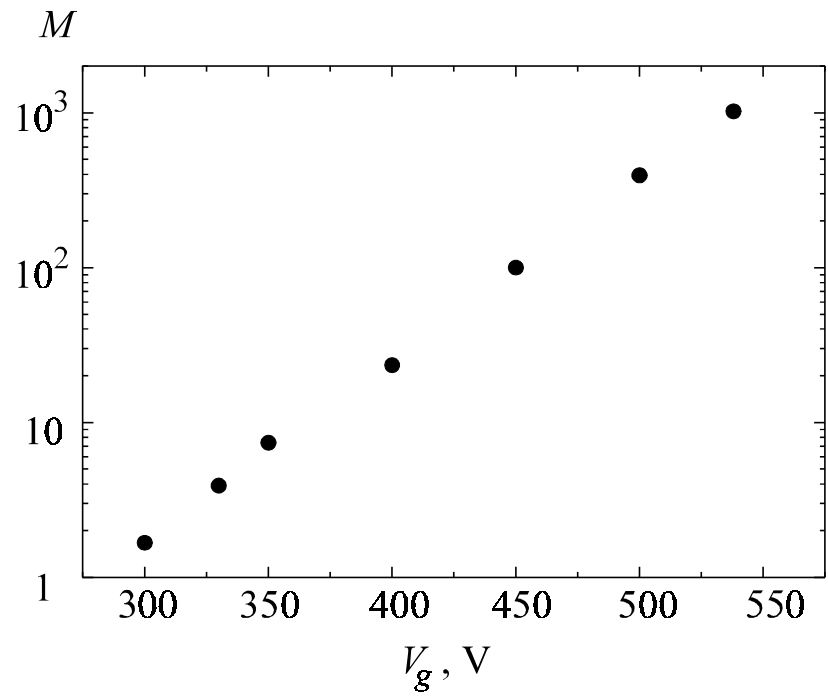

Fig. 5

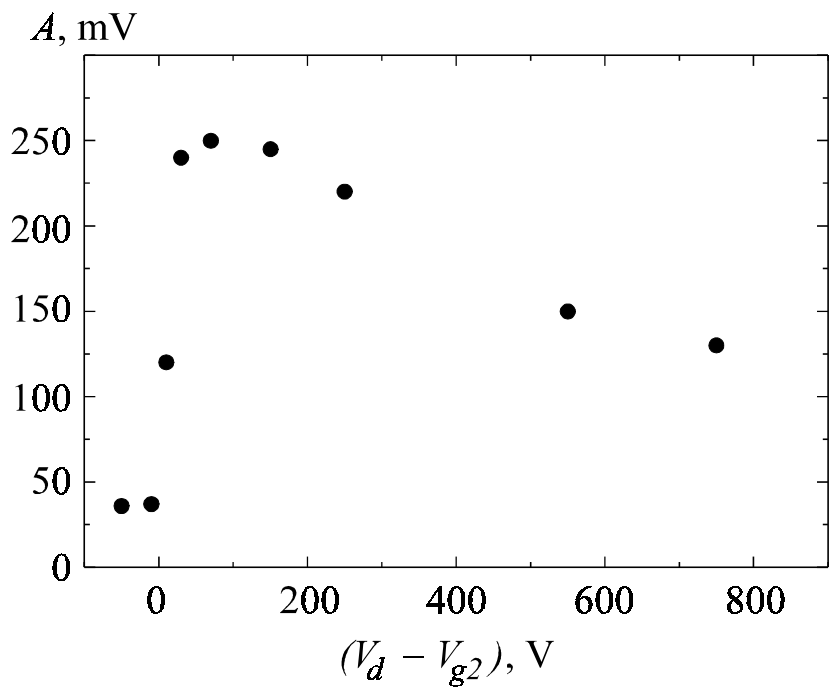

Fig. 6

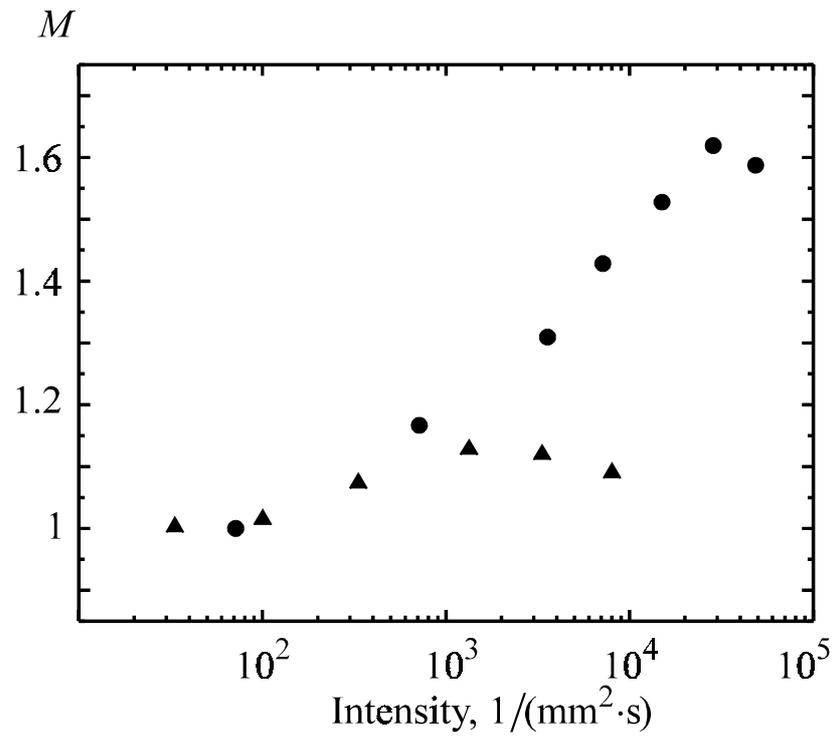

Fig. 7

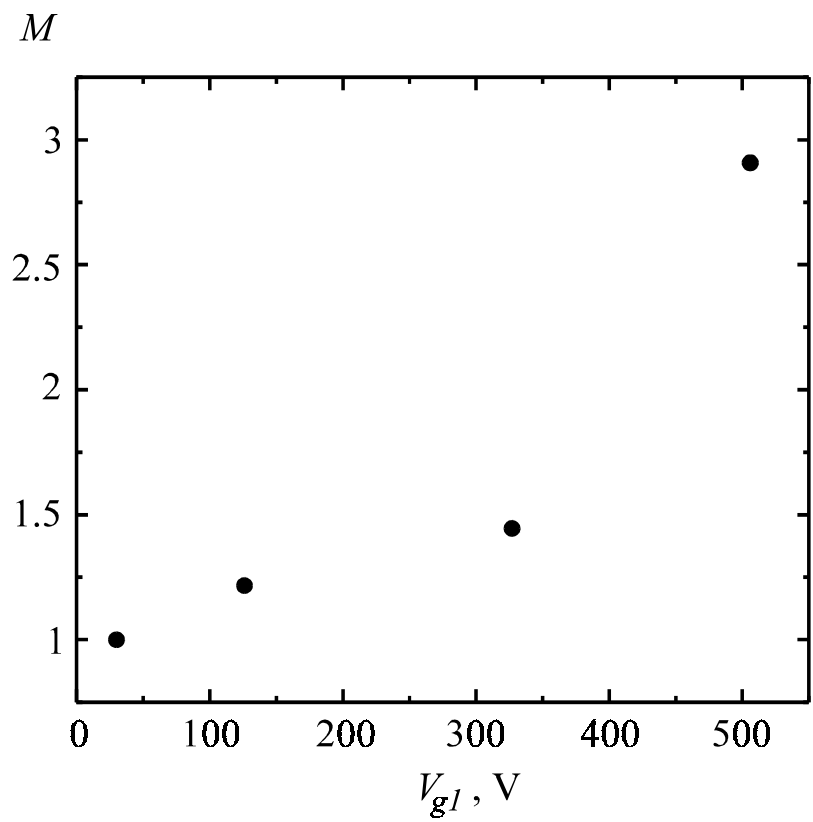

Fig. 8

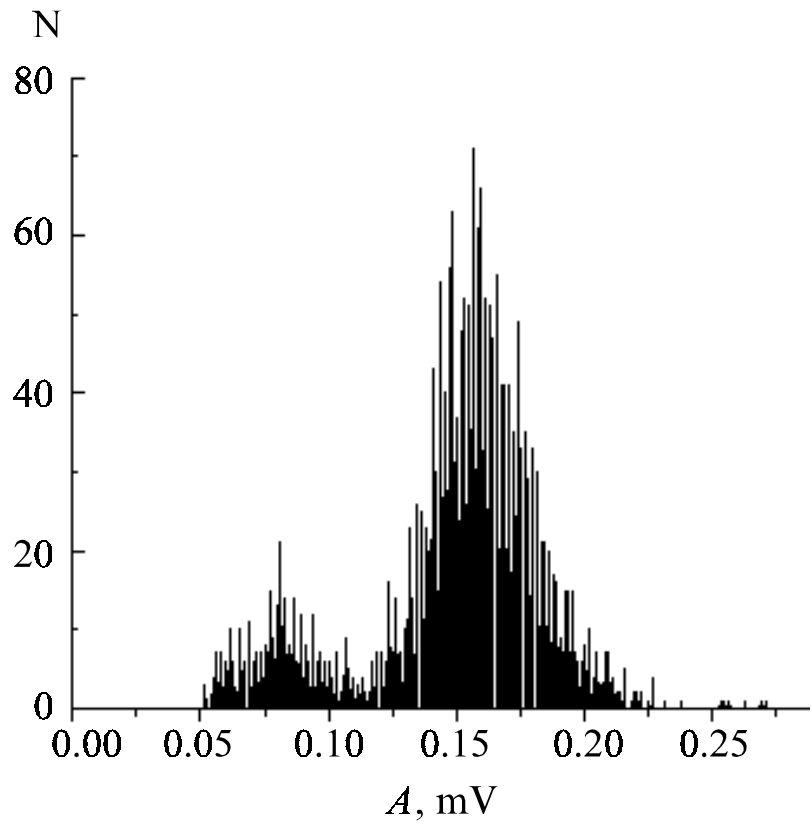

Fig. 9